\documentclass[journal=jacsat,manuscript=article,a4paper,layout=manuscript]{achemso}
\setkeys{acs}{articletitle = true,maxauthors = 0}

\usepackage{amsmath}
\usepackage[]{graphicx}
\usepackage{mathptmx}
\usepackage{helvet}
\usepackage{changebar}
\newcommand{\fig}[1]{Fig.~\ref{#1}}

\title{
High-Resolution Nuclear Magnetic Resonance Spectroscopy With
Picomole Sensitivity by Hyperpolarisation On A Chip}
\author{James Eills}
\altaffiliation{These authors contributed equally to this work}
\author{William Hale}
\altaffiliation{These authors contributed equally to this work}
\author{Manvendra Sharma}
\author{Matheus Rossetto}
\altaffiliation{Present address: University of York, United Kingdom}
\author{Malcolm H. Levitt}
\author{Marcel Utz}
\email{marcel.utz@soton.ac.uk}
\affiliation{School of Chemistry, University of Southampton, United Kingdom}
\date{\today}

\graphicspath{{./}}

\begin{document}

%\linenumbers
\nochangebars

\begin{abstract}
We show that high-resolution NMR can reach picomole sensitivity
for micromolar concentrations of analyte
by combining  parahydrogen induced hyperpolarisation (PHIP)
with a high-sensitivity transmission line micro-detector.
The para-enriched hydrogen gas is introduced into solution by diffusion
through a membrane integrated into a microfluidic chip.
NMR microdetectors, operating with sample volumes of a few $\mu$L or less,
benefit from a favourable scaling of mass sensitivity. However,
the small volumes make it very difficult to detect species present at less
than millimolar concentrations in microfluidic NMR systems.
In view of
overcoming this limitation, we implement parahydrogen-induced polarisation
(PHIP) on a microfluidic device with 2.5~$\mathrm{\mu L}$ detection volume.
Integrating the hydrogenation reaction into the chip minimises polarisation
losses to spin-lattice relaxation, allowing the detection of picomoles of
substance. This corresponds to a concentration limit of detection of better than
$\mathrm{1\,\mu M\,\sqrt{s}}$, unprecedented at this sample volume.
\cbstart
The stability and sensitivity of
the system allows quantitative characterisation of the signal dependence
on flow rates and other reaction parameters and permits homo-
(\textsuperscript{1}H-\textsuperscript{1}H) and heteronuclear
(\textsuperscript{1}H-\textsuperscript{13}C) 2D NMR experiments
 at natural \textsuperscript{13}C abundance.
\cbend
\end{abstract}

\maketitle

\begin{figure*}
  \cbstart
  \centering
  \includegraphics[width=15cm]{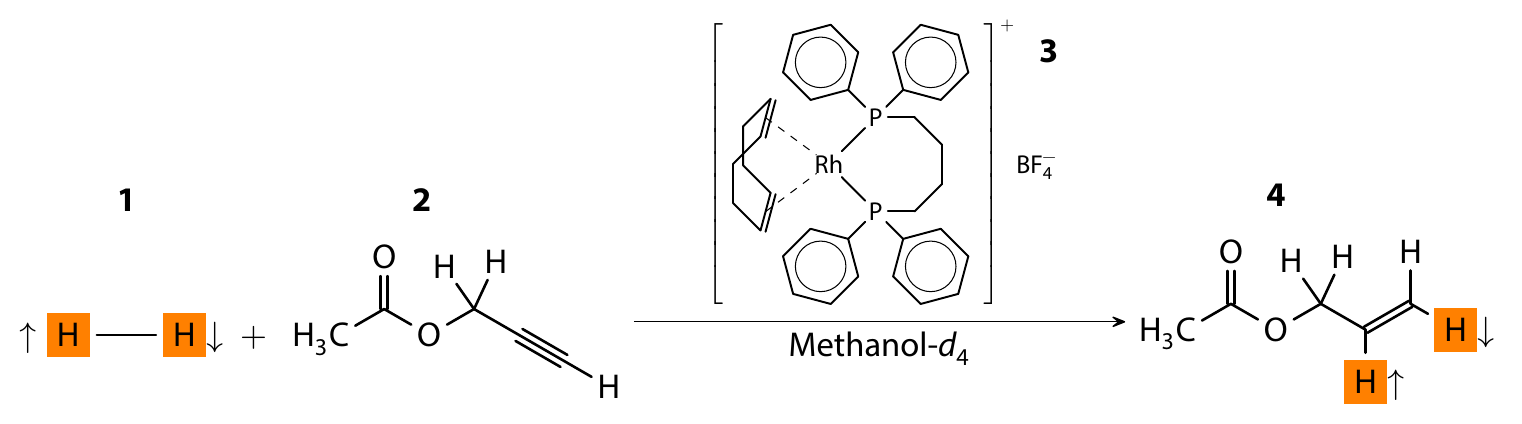}
  \caption{
    Scheme of the reaction used in the PHIP@chip experiment. Hydrogen gas
    \textbf{1}
    enriched in parahydrogen reacts with propargyl acetate \textbf{2} in
    the presence of the Rh catalyst \textbf{3} to form allyl acetate \textbf{4}.
  }
  \label{fig:reaction-scheme}
  \cbend
\end{figure*}

\begin{figure}
  \cbstart
	\centering
	\includegraphics[width=7cm]{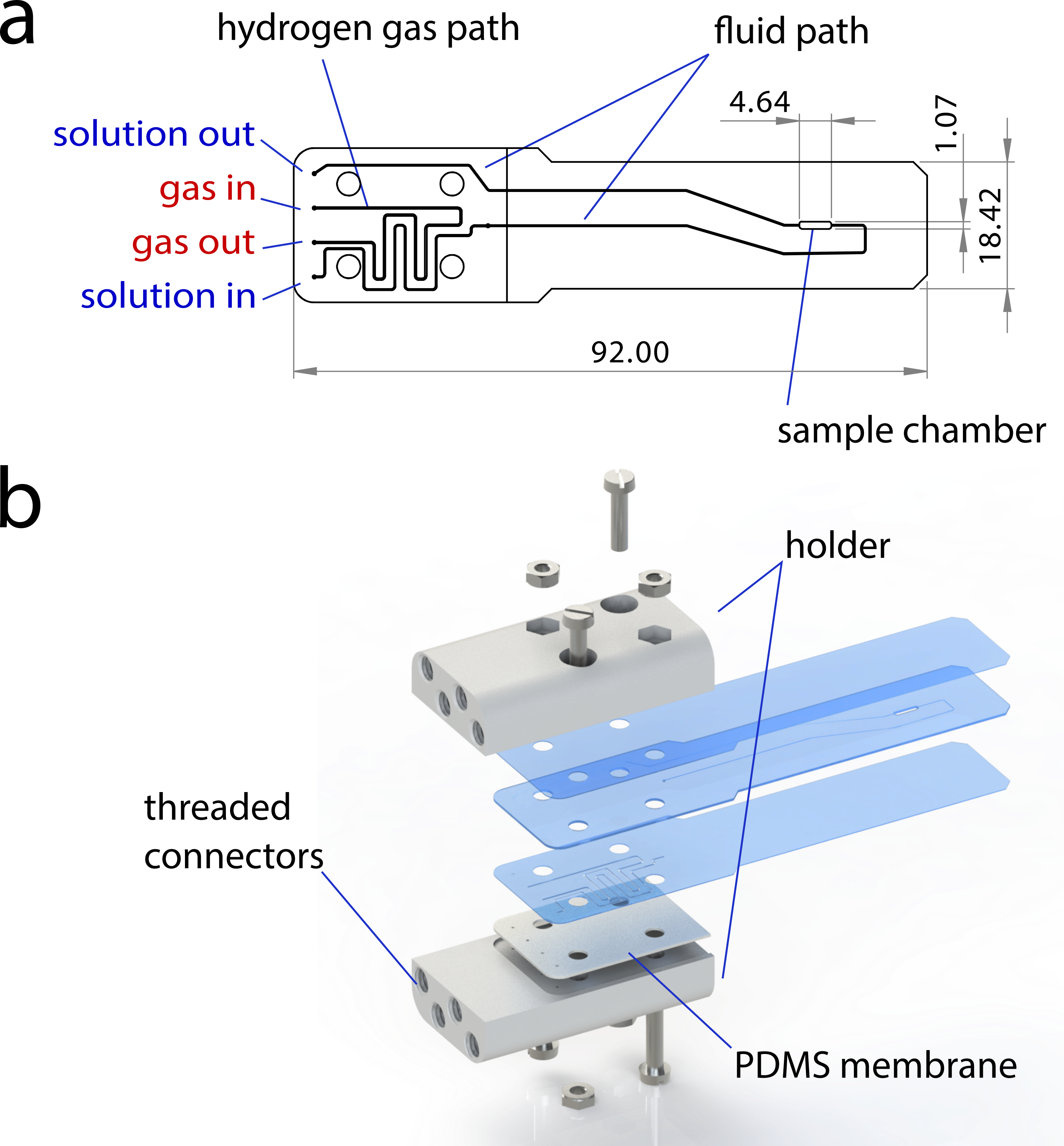}
	\caption{Overview of the PHIP@chip device.
		% a: scheme of the hydrogenation reaction;
    %
    a: outline drawing of the chip (dimensions in mm).
		b: CAD rendering of the chip assembly with individual chip layers
		separated, consisting of the PMMA chip, PDMS membrane, and two 3D
		printed holders with threads for the gas and fluid connections.
    The hydrogen gas
		diffuses through the PDMS membrane into the flowing liquid.
		}
	\label{fig:phip@chip1}
  \cbend
\end{figure}

\section{Introduction}
High-resolution NMR
spectroscopy is a superbly versatile method which provides detailed and
quantitative information on chemical composition and structure. It is widely
used to follow the progress of chemical reactions
\cite{Foley-quantitative:2004bk,*Foley:2014kpa,*Oosthoek-deVries:2019:ir},
as well as
metabolic processes in living systems
\cite{Wishart:2008ga,*Gottschalk:2008ixa,*CuperlovicCulf:2010vc,*shintuMetabolomicsonaChipPredictiveSystems2012a}.
However, NMR suffers from inherently low sensitivity,
which is due in part to the very weak polarisation of nuclear spins along the magnetic
field for samples in thermal equilibrium at ambient conditions.
Conventional high-resolution NMR therefore requires nanomole quantities of
sample. Many important problems require detection of analytes at low
micromolar concentrations, such as transient reaction intermediates, or
metabolic species. Despite the
comparatively higher mass sensitivity of NMR for small sample volumes
\cite{Olson:1995vu,Bart:2009kc}, conventional micro-NMR systems  around
1~$\mathrm{\mu L}$ achieve mass limits of detection of no better than\cite{Finch:2016gv}
1~nmol$\sqrt{\mathrm{s}}$,  corresponding to a concentration
limit of detection of 1~$\mathrm{m M \, \sqrt{s}}$. An increase of several
orders of magnitude in sensitivity is therefore required to enable NMR
studies of mass-limited samples at micromolar concentrations.

Microfluidic lab-on-a-chip devices are finding increasing applications in
chemistry and the
life sciences. They provide detailed control over the experimental
conditions at a much smaller length scale than conventional reactors, and
allow integration of synthesis, separation, and analytical steps on
a single platform \cite{Wang:2006en,*Theberge:2012iq,*Hoang:2011du,*Ohno:2008da,*Zhou:2004id,
*Fang:2018ib,*Hoang:2011ee,*Gunther:2006vd}. The small size also affords the possibility of
high experimental throughput.
In the life sciences, microfluidic devices are increasingly used as
sophisticated culture platforms for cells,
cell assemblies, tissues, and small organisms
\cite{Manz:1990vc,*Whitesides:2006vi,
*ElAli:2006ci,*West:2008jd,*Neuzil:2012gc,*Gracz:2015co}.
The integration of NMR with microfluidics
\cite{Ryan:2012ke,Badilita:2011td,Spengler:2014ir,Finch:2016gv,sharmaModularTransmissionLine2019} is promising, as
it enables in-situ, non-invasive monitoring of chemical and metabolic processes
in lab-on-a-chip systems.

The usefulness of microfluidic NMR could therefore be significantly
enhanced if the following conditions could be met:
(i) sample volumes around 1~$\mu$L or less;
(ii) a concentration limit of
detection near  1~$\mathrm{\mu M\,\sqrt{s}}$; and
(iii)  spectral
resolution of better than 0.01~ppm to allow distinction and identification
of chemical species.

Although exquisitely sensitive NMR detection schemes exist at low
magnetic fields,
approaching even single-spin detection in favourable cases
\cite{Rugar:1992dm,*Rugar:2004bc,*Mamin:2007ff,*Poggio:2010jf,
*Maze:2008cs,*Staudacher:2013kn,*Rugar:2015by,*McDermott:2002hp,
*Budker:2007hz,*Xu:2006kg,*Blanchard:2013gs}, they lack spectral resolution.
While a recent study
has demonstrated resolution of $J$ couplings using a nitrogen-vacancy
(NV) centre magnetometer\cite{Glenn:2018ct}, none of these
alternative detection schemes are compatible with high (several Tesla)
magnetic fields, which are essential to produce spectral dispersion
by chemical shifts.
So far, no method has been demonstrated with the combination of high
spectral resolution, high chemical dispersion,
and high sensitivity for small volumes required for
advanced microfluidic NMR measurements significantly
below the 1~mM concentration scale.

\cbstart
Hyperpolarisation methods generate substances which exhibit a transiently high
level of nuclear spin polarisation, much larger than thermal equilibrium.
As a result, hyperpolarised substances can give NMR signals that are enhanced
by more than 4 orders of magnitude \cite{munnemann2011nuclear}.
Hyperpolarisation methods can be combined
\cbend
with micro-NMR detectors and microfluidic systems \cite{McDonnell:2005dn,Desvaux:2009bq,Telkki:2010vg,Paciok:2011ek,JimenezMartinez:2014et,Causier:2015fg,eills-hale2018EuromarPHIP,Bordonali:2019jq}.
One such method
involves the chemical reaction of the singlet spin isomer of molecular
hydrogen, and is called parahydrogen-induced hyperpolarisation (PHIP)
\cite{hovener2018parahydrogen,duckett2012application,gloggler2013hydrogen,green2012theory}.

While most studies have so far brought the reaction liquid in direct contact
with hydrogen gas either through bubbling or by atomisation of the liquid
in a hydrogen-filled chamber \cite{bhattacharya2007towards,chekmenev2008pasadena,
chekmenev2009hyperpolarized,shchepin2014parahydrogen,
Reineri:2015he,cavallari201813,eills2017singlet},\cbstart
liquid-gas interfaces (and bubbles in particular)
\cbend
pose difficulties in the context of microfluidic devices, since they tend
to alter the flow properties, and can block fluid transport altogether.
Continuous delivery of parahydrogen by diffusion through
gas-permeable membranes has been demonstrated at
conventional size scales \cite{Roth:2010hk,Lehmkuhl:2018cd}.
It has been shown that silicone elastomer membranes can be used
to deliver parahydrogen directly to a flowing liquid in a microfluidic
device \cite{eills-hale2018EuromarPHIP}. Bordonali et al\cite{Bordonali:2019jq} have
recently combined a microfluidic NMR probe system with a gas exchange
chip based on a silicone elastomer membrane to implement the SABRE (signal
enhancement by reversible exchange) variant of parahydrogen-induced polarisation,
but achieved only small signal enhancement factors (3 to 4).

In distinction from previous work
\cite{bhattacharya2007towards,chekmenev2008pasadena,
chekmenev2009hyperpolarized,shchepin2014parahydrogen,
Reineri:2015he,cavallari201813,eills2017singlet,Lehmkuhl:2018cd},
we integrate the hydrogenation reactor into the chip itself, which greatly
reduces the polarisation losses due to spin-lattice relaxation.
As shown below, we achieve a signal enhancement factor over thermal polarisation
of about 1800, allowing detection of a
picomole quantity of analyte in a sample volume of $\mathrm{2.5\,\mu L}$,
while maintaining the full resolution of conventional $\mathrm{^1H}$
NMR spectroscopy.

This is accomplished by letting the parahydrogen gas diffuse through a
silicone elastomer membrane \cite{Lehmkuhl:2018cd}
to come into contact with a solution
flowing through the chip at a constant rate. The solution
contains a precursor, which is hydrogenated through a homogeneous
catalyst also present in the solution.
The microfluidic device is held in the bore of a conventional
NMR magnet using a purpose-built transmission line NMR probe.
This yields a continuous on-chip stream of hyperpolarised material. As shown
in the following, in addition to very high detection
sensitivities, this also results in a continuous and highly stable operation
of the system, making it possible to perform hyperpolarised
two-dimensional NMR experiments \cite{Roth:2010hk,Giraudeau:2009fn,Lloyd:2012cf,Eshuis:2015ce}.
By replacing the hyperpolarised gas feed with hydrogen gas at thermal
equilibrium, it is possible to gain kinetic information on the hydrogenation
process, as well as to calibrate the intensity of the hyperpolarised NMR signals.
This allows accurate assessment of the achieved polarisation levels, something
that has been notoriously difficult in the context of
parahydrogen-induced polarisation.

\section{Experimental}
The
microfluidic chips were constructed from three layers of cell cast
poly(methyl methacrylate) (PMMA) sheet
material (Weatherall Equipment).  The sheet thickness was 200 $\mu$m for the
top and bottom layers, and 500 $\mu$m for the middle layer. The fluid and gas
channels were designed on AutoCAD and cut into the PMMA using a laser cutter
(HPC Laser L3040) to a width and depth of 150 $\mu$m. The layers were
subsequently bonded together with a plasticiser (2.5\% v/v dibutyl phthalate in
isopropyl alcohol) under heat and pressure (358 K, 3.5
tonnes) \cite{Yilmaz:2016fx}. The total internal fluid volume is 4~$\mu$L, and
the sample chamber is 2.5~$\mu$L.

The device also employs a poly(dimethyl siloxane) (PDMS) membrane
(Shielding Solutions) to facilitate para-H\textsubscript{2} transport,
of 1 mm thickness with laser-cut screw holes. The parahydrogen
polarisation lifetime in the PDMS after O\textsubscript{2} removal was
measured to be \textasciitilde{}4 h (see supplementary information). The
PMMA chip and PDMS membrane layer are sealed with a pair of
screw-tightened 3D printed (Accura Xtreme, Proto Labs) holders, with
fluid and gas in/out ports (to fit Kinesis UK NanoPorts).

The assembled microfluidic device was put in a transmission line based
home-built probe \cite{sharmaModularTransmissionLine2019}. The
device sits between the two detector planes with the
sample chamber of the device aligned with the sensitive area of the
detector. All NMR experiments were performed at a field strength
of 11.7~T with an AVANCE III console. Nutation frequencies for RF pulses
were 100~kHz for protons, and 20~kHz for carbon in the case of the HMQC
spectrum. 16k data points were acquired over 1.2~s for proton 1D spectra.
Saturation recovery experiments used a train of 512 $\pi/2$ pulses
separated by a delay of 0.1~ms, followed by a recovery delay, and a $\pi/4$
excitation pulse.
The PH-TOCSY spectrum was acquired using the States-TPPI method,
with 256 $t_1$ increments, averaging 8 transients per increment.
2048 complex data points in 0.2~s were acquired for each increment.
The PH-HMQC experiment was acquired using the States method, with
128 $t_1$ increments, averaging 8 transients with 2048 complex points
over 0.2~s. 1D spectra were processed using MestreNova (Mestrelab, Italy).
2D spectra were processed using scripts written in Julia \cite{Bezanson:2017gd}.

To generate parahydrogen gas at 50\% para enrichment, hydrogen gas
(purity 99.995\%) was passed through a home-built parahydrogen generator
containing an iron (III) oxide catalyst cooled to 77~K using liquid
nitrogen.

The solution before reaction contained 20 mM propargyl acetate \textbf{2}
and 5 mM
1,4-bis(diphenyl\-phosphino)\-butane(1,5-cyclo\-octadiene)\-rhodium
tetra\-fluoro\-borate \textbf{3} in methanol-d\textsubscript{4}. In an attempt to
avoid possible spin relaxation or chemical side-reaction effects,
dissolved oxygen from the atmosphere was removed by 5 minutes of
vigorous helium bubbling.

The parahydrogen gas was delivered through a PTFE tube (1/16 inch O.D.,
1/32 inch I.D.) into the 3D printed chip holder, and out via a second
PTFE line, using a mass flow controller (Cole-Parmer) to limit the flow
to 20 mL min\textsuperscript{-1} at an overpressure of 5 bar. Although
most of the parahydrogen gas passes directly through the system, some
amount dissolves into the PDMS layer, which in terms of
H\textsubscript{2} solubility behaves similarly to other organic
solvents. The solution was loaded into a 3.5 mL plastic syringe with a
Luer lock connection to in-flow PEEK tubing (1/16 inch O.D., 0.007 inch
I.D.) leading to the chip. The same tubing was used for the solution
out-flow into a container exposed to a back pressure of 1.5 bar of
nitrogen gas, to prevent formation of hydrogen bubbles in the chip.
Solution flow into the chip was controlled with a syringe pump (Cole-Parmer).

\section{Results and Discussion}
The hydrogenation reaction system employed in the present work is shown in
\fig{fig:reaction-scheme}.
Para\-hydrogen-en\-riched hydrogen gas $\mathbf{1}$ was
allowed to react with propargyl acetate $\mathbf{2}$, in the presence of
a rhodium catalyst $\mathbf{3}$. The substrate $\mathbf{2}$ was chosen in
view of future studies based on side-arm hydrogenation
\cite{Reineri:2015he,cavallari201813,cavallari2015effects}.

\fig{fig:phip@chip1} \cbdelete shows the microfluidic device used for the
present study. It consists of a chip made from PMMA, which houses a sample
chamber of 2.5~$\mathrm{\mu L}$ volume that aligns with the transmission line
detector of a home-built NMR probe assembly, which was fitted inside of an
11.7~T NMR magnet.
\cbstart Fluid is flowed through the chip by means of a syringe pump
\cbend
installed outside of the magnet bore; connections are made through threaded
ports in the two 3D-printed holders shown in \fig{fig:phip@chip1}b.
Para-enriched H\textsubscript{2} gas at 5~bar above ambient pressure flows
through a second
channel in the chip\cbdelete , which runs in
the immediate vicinity of the liquid channel
(a depiction of the set-up is given in the SI).

\begin{figure}
  \cbstart
	\centering
	\includegraphics[width=8cm]{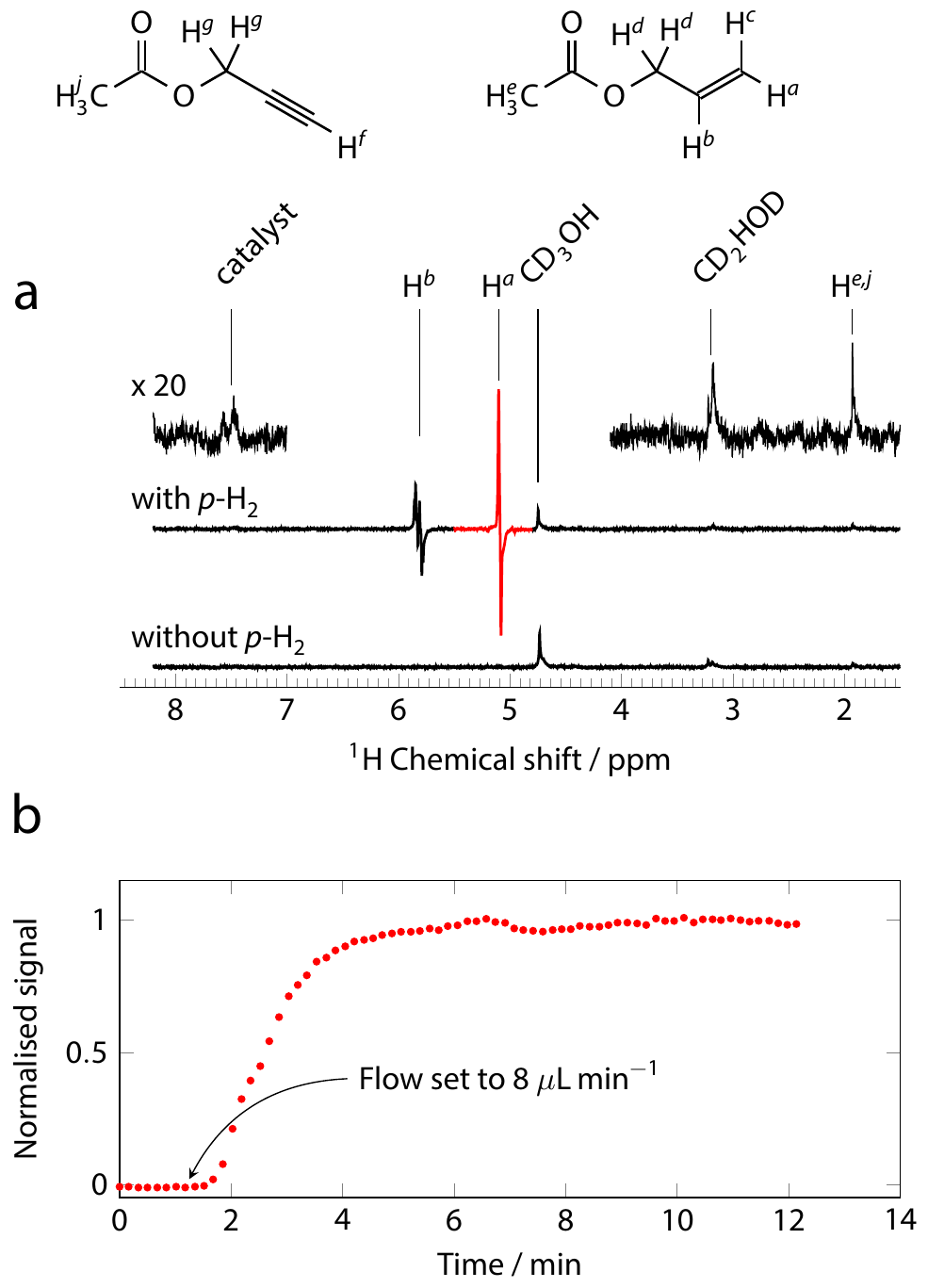}
	\caption{a: Single-scan proton NMR spectrum obtained with
  parahydrogen at 5 bar using the PHIP@chip setup at a continuous
  flow rate of 8~$\mu$L$\,\text{min}^{-1}$ (top trace with enlargement).
  Antiphase doublets from the two
	hyperpolarised protons $\mathrm{H}^a$  and $\mathrm{H}^{b}$ are
  clearly visible at 5.2~ppm and 5.9~ppm, respectively. Without
  parahydrogen, these signals are not observed (bottom trace).
	b: Buildup of the hyperpolarised signal ($\mathrm{H}^{a}$)
  after initiation of flow.
	}
	\label{fig:phip@chip2}
  \cbend
\end{figure}

The chip consists of three laser-cut layers of poly methylmethacrylate (PMMA)
bonded together, as shown in \fig{fig:phip@chip1}b.\cbdelete\ Channels in the
left part of the chip, where it is clamped between the holders, are cut through
the top layer, while they are scored into the middle layer of the chip (and
hence sealed from the outside) in the free part of the device.
Within the clamps, the exposed channels are sealed by means of a PDMS membrane.
The flowing liquid as well as
the pressurised hydrogen gas are therefore exposed to the PDMS layer,
which serves as a diffusion bridge for the hydrogen.
The holders, made by 3D printing, keep the membrane and the chip aligned,
and maintain mechanical pressure to ensure sealing. Channels
inside the holders guide the fluid and gas to and from
the four access points at the top end
of the chip, as shown in \fig{fig:phip@chip1}b.
The PDMS membrane acts both as a diffusion conduit for hydrogen gas
and as a fluid seal.
In a crucial difference to the otherwise similar geometry of the
hydrogenation chip used by Bordonali et al\cite{Bordonali:2019jq},
the gas and liquid channels are arranged side by side, and molecular
\cbstart
hydrogen diffuses through the bulk of the PDMS material rather than across
the membrane.
\cbend
Clamping the PDMS membrane onto the chip using the
holders, this makes it possible to use large gas pressures (up to 5 bar in
the present experiments). This would be difficult to achieve
\cbdelete if the liquid and gas channels were
arranged on opposite sides of the membrane.

\cbstart
\fig{fig:phip@chip2}a shows a single-scan proton NMR spectrum
obtained from a steady-state PHIP@chip experiment (top trace), compared
to the spectrum obtained without parahydrogen (bottom trace).
The hyperpolarised spectrum is dominated by an antiphase doublet,
centred at 5.17~ppm, and an antiphase multiplet at 5.92~ppm, corresponding to
protons in the $\mathrm{H}^a$ and $\mathrm{H}^b$ positions of the
hydrogenation product \textbf{4}.
The PDMS membrane is equilibrated with para-enriched hydrogen gas, which is
supplied from an aluminium storage tank at a regulated pressure of 5~bar. The
gas flow rate is kept constant at 20~$\text{mL}\,\text{min}^{-1}$ by means of a
mass flow controller placed after the chip. This ensures that the gas channel
always contains fresh para-enriched hydrogen gas at the design pressure of
5~bar. The fluid channel of the chip is pre-filled with a solution of 20~mM
precursor $\mathbf{2}$ and 5~mM catalyst $\mathbf{3}$ in methanol-$d_4$. NMR
spectra are acquired every 30~s, using a $\pi/4$ excitation pulse.  The fluid
channel is connected to a syringe pump situated outside the NMR magnet.
The
liquid flow is started by setting the target flow rate on the syringe pump to
8~$\mathrm{\mu L\,\text{min}^{-1}}$ (marked by an arrow \fig{fig:phip@chip2}b).
The NMR signal intensity begins to rise about 30~s later, and reaches a steady
state after about two minutes.
\cbend

\begin{figure}
  \cbstart
	\includegraphics[width=7cm]{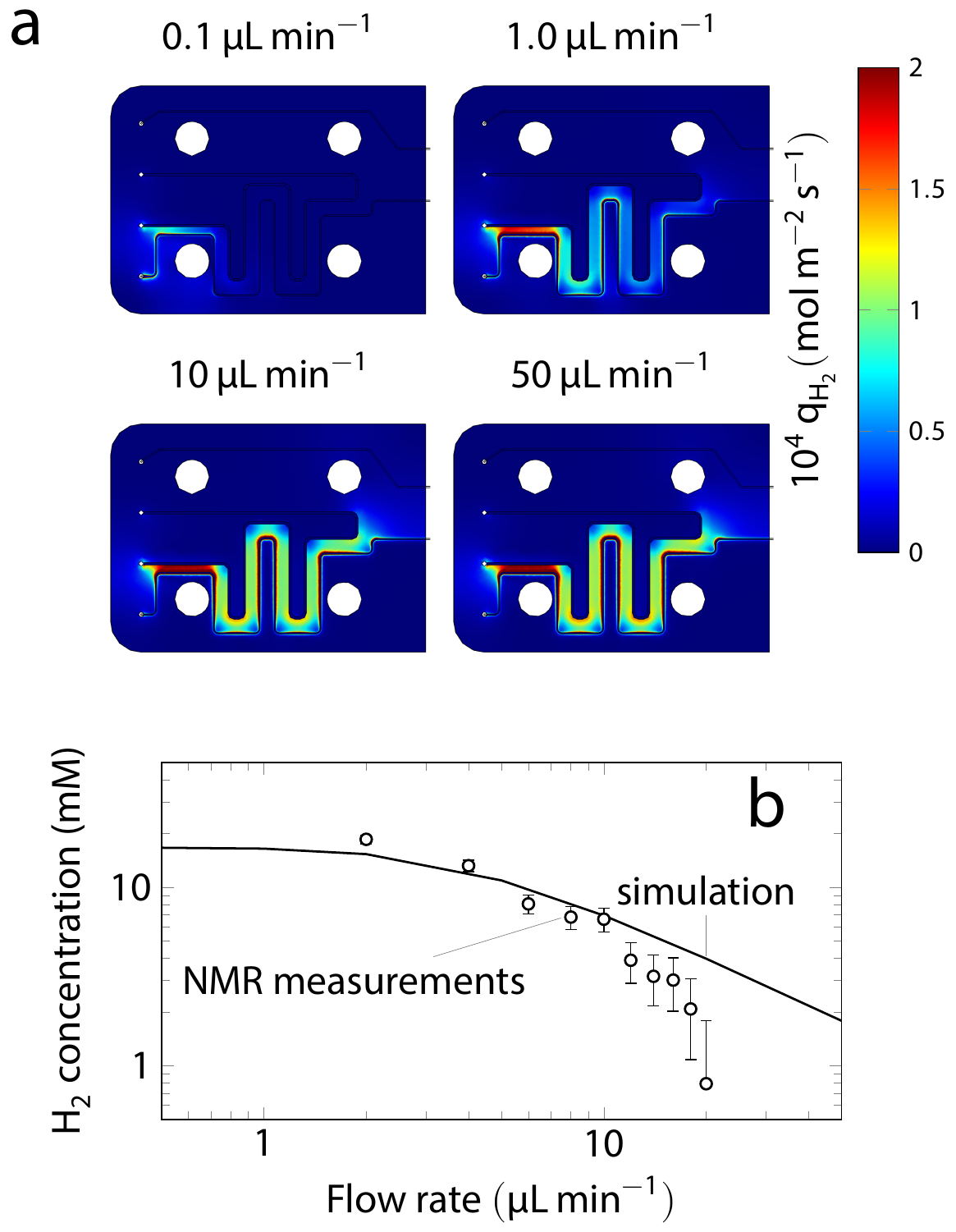}
	\caption{
		Finite element simulation of hydrogen uptake. a: Diffusive hydrogen
		flux in the PDMS membrane for different liquid flow rates;
		b: final hydrogen concentration in flowing methanol as a function of
		flow rate. Solid line: simulation, open circles: NMR measurements.
	}
	\label{fig:h2fluxsim}
  \cbend
\end{figure}

The hydrogen transport through the membrane and its uptake into the flowing
liquid was simulated using two coupled finite element models: a dilute species
diffusion model for hydrogen gas in the PDMS membrane, and a dilute species
diffusion and convection model for hydrogen dissolved in the flowing liquid. The
hydrogen partial pressures at the liquid/PMDS interface are constrained to be
equal, and the hydrogen partial pressure at the gas/PDMS interface was set to a
fixed value of 5~bar. \fig{fig:h2fluxsim}a shows the diffusive flux of hydrogen
through the PDMS membrane.  Since the gas/PDMS interface acts as a source, and
the liquid/PDMS interface as a sink for hydrogen, the flux is strongest where
the two channels are in close proximity. At the lowest flow rate, significant
transport only takes place in a very small area, and the liquid is saturated
with hydrogen within the first few mm of the path which is in contact with the
PDMS. The higher the flow rate, the further the area of significant flux extends
downstream. At about 10~$\mathrm{\mu L \,min^{-1}}$, the hydrogen flux covers
the entire length of the area between the liquid and gas channel interfaces. The
finite element model also predicts the resulting concentration of hydrogen in
the liquid (methanol) as a function of flow rate. This is shown by the solid
line in  \fig{fig:h2fluxsim}b. The circles represent NMR measurements. At
flow rates between 2 and 10~$\mathrm{\mu L min^{-1}}$, experimental results are
in good agreement with the simulation. At higher flow rates, however, the
experimentally observed hydrogen concentrations are significantly lower than the
predictions. It is currently unclear what causes this discrepancy; possibly high
flow rates lead to deformation of the PDMS layer over the liquid channel and
thus change the uptake geometry. At flow rates below 10$\,\mathrm{\mu L
\,min^{-1}}$, the simulation and experiments both indicate that the flowing
solvent is nearly saturated with hydrogen. Detailed information on the finite
element simulations, as well as data on parahydrogen partial pressure
throughout the chip is given in the supporting information.

\begin{figure}
  \cbstart
	\centering
	\includegraphics[width=7cm]{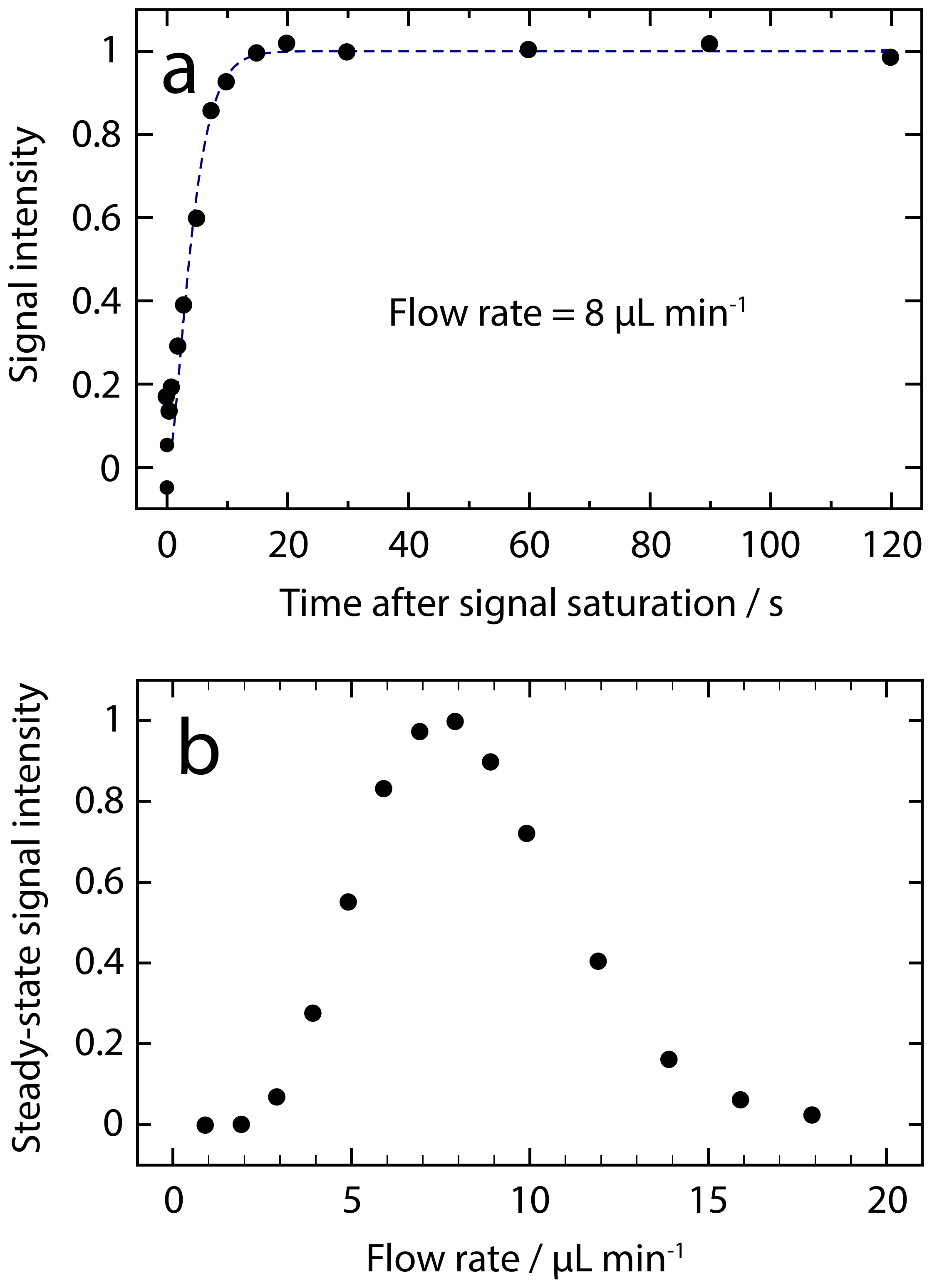}
	\caption{Saturation recovery results.
	a: Signal buildup at constant
	flow rate after saturation (solid dots: measured data points,
  the dashed line is a guide to the eye);
  %\cbend
	%
	b: Magnitude of the steady-state signal after full recovery (at least
	100~s after saturation) as a function of flow rate. A clear maximum
	at 8~$\mu\mathrm{L}\,\text{min}^{-1}$ is observed.
	}
	\label{fig:satrec-summary}
  \cbend
\end{figure}

Clearly, the steady-state signals observed at constant flow rate are the result
of a dynamic equilibrium between the rate of hydrogenation, the rate of
transport of the hydrogenated product to the sample chamber and its removal
from it, and spin-lattice relaxation. In order to probe the interplay of these
factors, the NMR signal was suppressed by saturating the spin populations
with a train of 512 $\pi/2$ pulses separated by 100 $\mu$s delays.
The signal intensity was then measured as a function of the delay between the
end of the saturation train and the NMR excitation pulse.
\fig{fig:satrec-summary}a shows an example of the data thus obtained at a
flow rate $q=8\,\mathrm{\mu L\,\text{min}^{-1}}$ (results for
other flow rates are given in the SI).
The signal increases rapidly after saturation, reaching
steady-state levels after about 10~s.

\begin{figure}
  \cbstart
  \begin{center}
    \includegraphics[width=8cm]{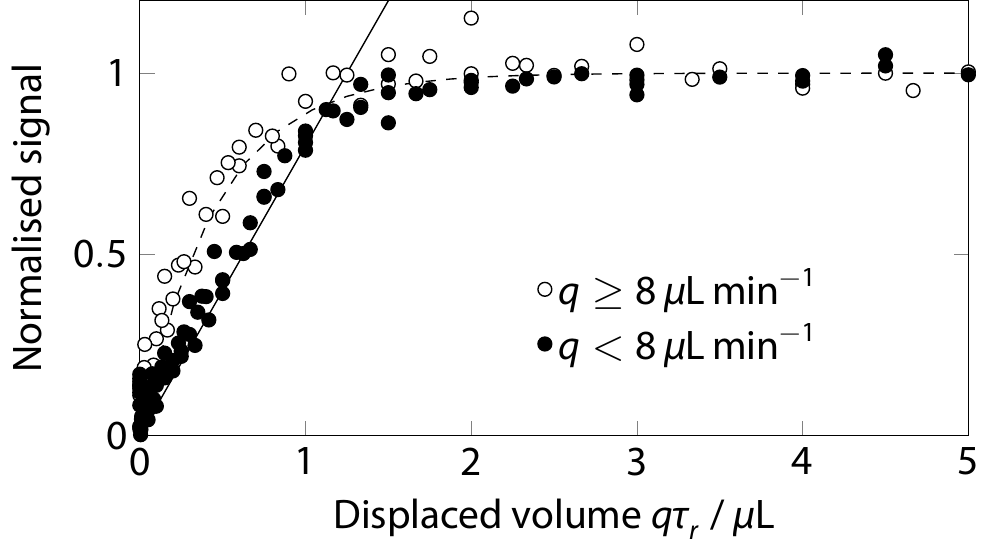}
  \end{center}
  \caption{Signal recovery after saturation, normalised by the maximum signal
  observed at long recovery times. The horizontal axis is the volume
  moved through the chip during the recovery time $\tau_r$, i.e., $q\,\tau_r$,
  where $q$ is the flow rate. Filled circles correspond to flow rates below
  the optimum ($q<8\,\mu\mathrm{L}\,\text{min}^{-1}$), where as open circles
  are obtained at flow rates $q\ge 8\,\mu\mathrm{L}\,\text{min}^{-1}$. The solid
  and dashed lines are guides to the eye for the solid and open circle data points,
  respectively.}
  \label{fig:displaced-volume}
  \cbend
\end{figure}

The intensity of the steady-state NMR signal exhibits a clear maximum with flow
rate (\fig{fig:satrec-summary}b), reflecting a balance between hydrogen uptake,
reaction kinetics, and spin-lattice relaxation. The optimum, with the largest signal at
saturation, is reached at a flow rate of 8~$\mathrm{\mu L\,\text{min}^{-1}}$.
The nature of the stationary state established in the system at each
flow rate becomes clearer if the saturation recovery data is plotted in terms
of the volume displaced during the saturation recovery time $q\tau$, rather
than the recovery time itself, and normalised to the steady-state signal intensity
at each flow rate, as shown in \fig{fig:displaced-volume}. At flow rates below
the intensity maximum at $q<8\,\mathrm{\mu L\,\text{min}^{-1}}$ (solid
circles), \cbdelete
the data points collapse onto a curve that shows an initial linear increase
up to a displaced volume of about 1~$\mu$L, followed by rapid saturation to the
steady-state value. This \cbdelete\ behaviour clearly indicates that the
signal recovery in this regime is dominated by the convective fluid transport.
At these flow rates, a constant concentration of
hyperpolarised material is established in the flowing liquid upstream of the
sample chamber, and is simply carried back into view of the NMR detector
after the saturation pulses end. \cbdelete
The maximum signal is reached after a volume
of about 1.5~$\mu$L has been displaced. This is less than the capacity
of the sample chamber, reflecting the uneven velocity distribution inside it.
At flow rates above the optimum ($q\ge 8\,\mathrm{\mu L\,\text{min}^{-1}}$),
a somewhat different behaviour is observed. The initial recovery rate is
faster (\fig{fig:displaced-volume}, open circles), and appears to follow
an exponential rather than linear shape. This suggests that at these flow
rates, the stationary state is not yet established at the point where the
liquid enters the sample chamber, and therefore, the observed recovery
is dominated by the ongoing hydrogenation reaction. A detailed kinetic
analysis of these processes is beyond the present scope, but is underway
in our laboratory, and will be reported on a later occasion.

\begin{figure}
  \cbstart
  \begin{center}
    \includegraphics[width=6.0cm]{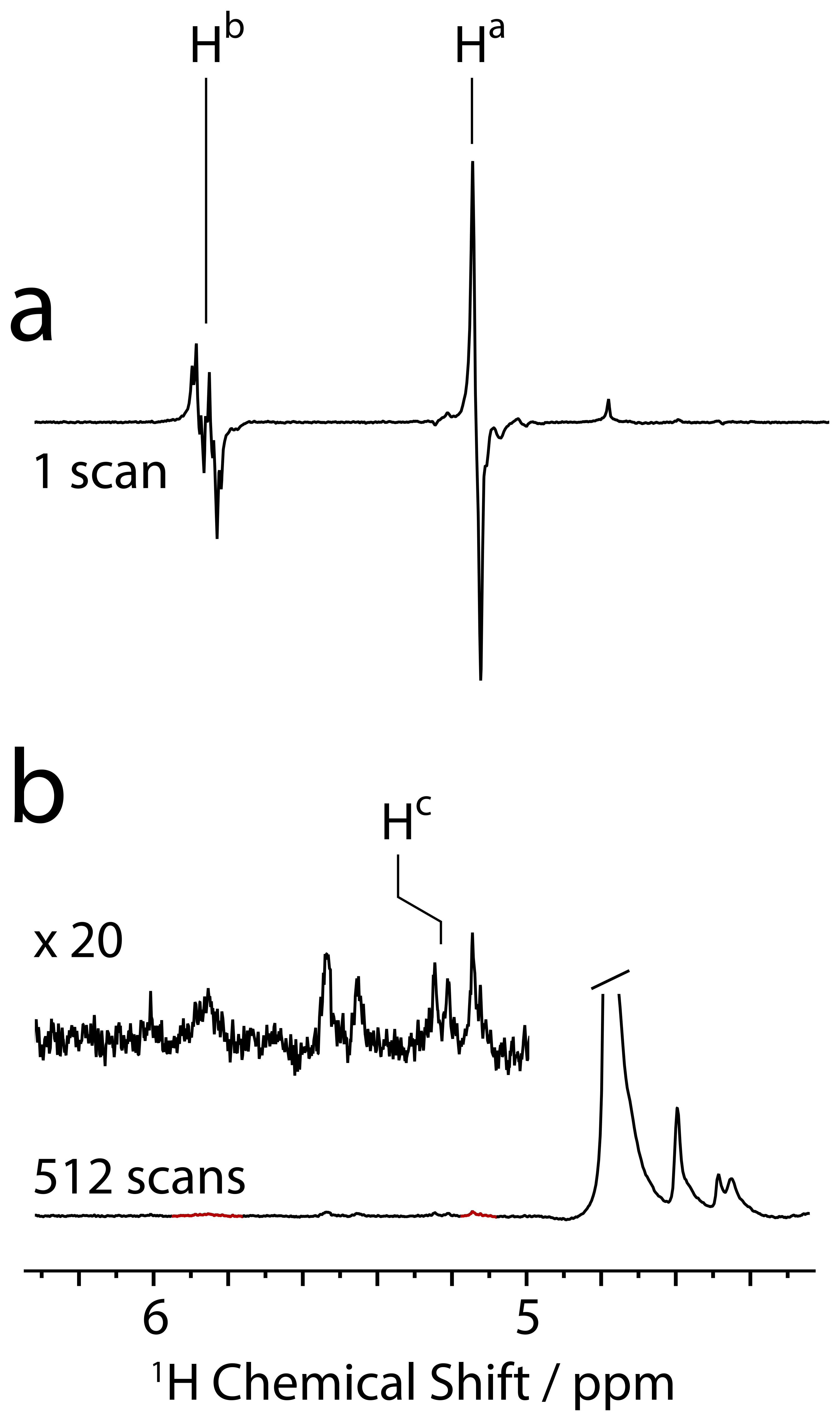}
  \end{center}
  \caption{	a: Single-scan steady-state spectrum obtained at the optimum flow rate
  	with para-enriched H\textsubscript{2}; b: spectrum obtained at the same flow
    rate with hydrogen gas
  	in thermal equilibrium. 512 transients have been averaged. Signal enhancement by
  	PHIP was determined by comparing the integral of the positive lobe of
    the $\mathrm{H}^a$ signal in spectrum a to the
  	integral of the corresponding (purely absorptive) peak in spectrum b. A fully labelled spectrum is
  	is given in the SI.}
  \label{fig:pH2-vs-thermal512}
  \cbend
\end{figure}

In order to determine the sensitivity of detection of the hydrogenation product
at the optimum flow rate, the experiment was repeated using normal hydrogen.
In this case, the signal from  protons $\mathrm{H}^a$ and $\mathrm{H}^b$
of the hydrogenation product \textbf{4}
are too weak to be observed above the noise in a single scan.
\cbstart
\fig{fig:pH2-vs-thermal512}
compares the hyperpolarised signal (a) to the averaged signal of 512 transients
obtained with hydrogen in thermal equilibrium (b).

Since the methyl group in the
precursor and the hydrogenation product contribute to the same signal at 2.05~ppm
(signal labelled $\mathrm{H}^{e,j}$ in \fig{fig:phip@chip2}a),
\cbend
this signal can be used as a calibration standard, with a concentration of 20~mM
which is unaffected by the hydrogenation reaction. By comparing this integral to that
of the signal from the $\mathrm{H}^a$ protons, the concentration
of hydrogenated product can be
quantified. At a flow rate of 8~$\mathrm{\mu L\,\text{min}^{-1}}$, an allyl acetate
(product) concentration of $(0.29\pm 0.05)\,\mathrm{mM}$ was found, corresponding to a total
of $(0.725\pm0.125)\,\text{nmol}$ in the $2.5\,\mathrm{\mu L}$ sample volume.

This quantity can be used to determine the limit of detection of the
hyperpolarised product. The signal/noise ratio (SNR) in the spectrum shown in
\fig{fig:pH2-vs-thermal512}a is 400($\pm 10\%$), and the line width is $6\pm
0.5\,\text{Hz}$. The normalised limit of detection is given by
\begin{equation}
\text{nLOD}_\omega = \frac{3 n}{\text{SNR}\,\sqrt{\Delta f}},
\end{equation}
 where $n$ is
the amount of sample and $\Delta f$ is the signal bandwidth. In the present
case, one finds $\text{nLOD}_\omega = (2.2\pm
0.4)\,\text{pmol}\,\sqrt{\text{s}}$. Limits of detection in this range have so
far only been reported in very limited circumstances, including
chemically-induced dynamic nuclear polarisation (CIDNP)
\cite{mompean2018pushing}, or by making use of unconventional low-field
detection systems
such as force-detected magnetic resonance or optical detection methods\cite{Rugar:1992dm,*Rugar:2004bc,*Mamin:2007ff,*Poggio:2010jf,
*Maze:2008cs,*Staudacher:2013kn,*Rugar:2015by,*McDermott:2002hp,
*Budker:2007hz,*Xu:2006kg,*Blanchard:2013gs}. In the
present case, we are using conventional inductive detection, and retain the full
resolution and specificity that make \cbstart high-field \cbend
proton NMR spectroscopy such a powerful
analytical tool.

\begin{figure}
  \cbstart
	\centering
	\includegraphics[width=5cm]{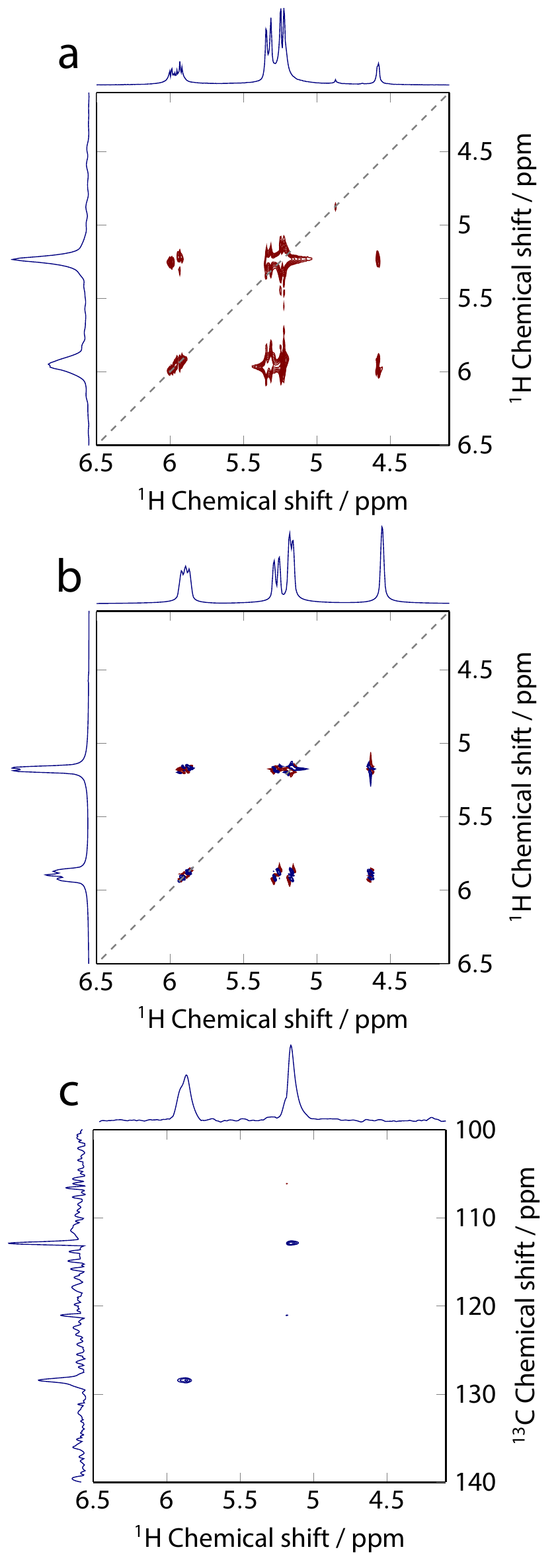}
	\caption{
		The continuous flow PHIP@chip approach allows acquisition of
		two-dimensional spectra with very high sensitivity.
    a: PH-TOCSY spectrum of the hyperpolarised reaction mixture,
		flowing at 8~$\mu\mathrm{L}\,\text{min}^{-1}$.
    b: Simulated PH-TOCSY spectrum. The diagonal in the spectrum is marked by a dashed grey
		line. Only the protons originating from parahydrogen give signals on
		the diagonal; the polarisation is transferred to the other locations by
		the isotropic mixing sequence. Both PH-TOCSY spectra are plotted in
    magnitude mode (phase sensitive spectra are given in the SI).
    c: $^1$H-$^{13}$C PH-HMQC spectrum
		showing two separate multiplets, each correlating one of the two
		hyperpolarised protons with the directly bonded $^{13}$C spin.
}
	\label{fig:PH-TOCSY-HMQC}
  \cbend
\end{figure}

The mass limit of detection (LOD) for
protons at a magnetic field of 14.1~T (corresponding to a proton Larmor frequency
of 600~MHz) in state-of-the-art commercial NMR probes with a
conventional sample volume of 0.5~mL is approximately
100~$\text{nmol}\,\sqrt{\text{s}}$.
Microfluidic NMR systems can make use of miniaturised NMR detectors,
which benefit from a favourable scaling of the mass sensitivity
with detection volume \cite{Olson:1995vu,Badilita:2011td,Zalesskiy:2014hi}. At a size scale
of 2.5~$\mu\mathrm{L}$, a mass sensitivity around
$1\;\text{nmol}\,\sqrt{\text{s}}$
has been reported \cite{Finch:2016gv}.
However, due to the limited volume in such systems,
the \emph{concentration} sensitivity is very poor, such that
only compounds present at mM levels can be quantified
in microfluidic NMR systems. This situation gets worse as the detector
volume decreases. By contrast, many samples of interest, such as
metabolites in microfluidic culture systems, are only present
at $\mu$M levels.
In the present
case, the concentration limit of detection is
\begin{equation}	\text{cLOD}_\omega =
\frac{\text{nLOD}_\omega}{V} = (0.88\pm0.16)\,\mathrm{\mu M\,\sqrt{s}}.
\end{equation}

From the ratio of the signal intensities in the thermal and hyperpolarised
spectra shown in \fig{fig:pH2-vs-thermal512}a and b, it is possible to estimate the
$\mathrm{^1H}$ polarisation levels. In the thermal spectrum, the SNR is about
5:1, whereas it is 400:1 in the hyperpolarised spectrum. The thermal spectrum is
obtained from 512 transients, therefore the single transient thermal SNR would
be $5/\sqrt{512}\approx 0.22$. This leads to a signal enhancement factor of
$\epsilon\approx 400/0.22 \approx 1800$.

This can be compared to the expected signal enhancement given the enrichment
level of para-hydrogen used in the experiment. The ideal enhancement factor is
given by
\cbstart
\begin{equation}
	\epsilon_{id} =\frac{4x_p-1}{3} \frac{1}{\sqrt{2}}\frac{k_BT}{\hbar \gamma B_0},
\end{equation}
where $x_p$ is the mole fraction of parahydrogen in the feed gas, $\gamma$ is
the magnetogyric ratio, $B_0$ is the magnetic field, and $\hbar$ and $k_B$ are
Planck's and Boltzmann's constants, respectively.
The factor $\frac{1}{\sqrt{2}}$ reflects the use of a $\pi/4$ pulse for the
hyperpolarised experiment. At a temperature of $T=298$~K
and a magnetic field of $11.7$~T, and with $x_p=0.5$, this yields
$\epsilon_{id}\approx 5900$, which is a factor of 3.3 larger than the
experimentally observed enhancement factor. We can therefore conclude that
about 2/3 of the theoretically available spin order is lost to relaxation under
the present experimental conditions.
\cbend

A great advantage of the continuously operating microfluidic PHIP system is the
ability to acquire many transients in succession under virtually unchanged
conditions.
This is difficult to achieve with bubbling hydrogen through a solution.
As a consequence, hyperpolarised multi-dimensional NMR spectra\cite{Mishkovsky:2008cl,Giraudeau:2009fn,Roth:2010hk,Lloyd:2012cf,Eshuis:2015ce,Kiryutin:2019hy} have been recorded
either using automated reactors combined with NMR flow probes,\cite{Lloyd:2012cf,Eshuis:2015ce}
or using ultrafast acquisition techniques\cite{Mishkovsky:2008cl,Giraudeau:2009fn,Kiryutin:2019hy}.

The PHIP@chip setup allows straightforward acquisition of 2D spectra, using
conventional $t_1$ incrementation.
To demonstrate this, we have taken 2D TOCSY (Total Correlation
Spectroscopy) and HMQC (Heteronuclear Multiple Quantum Coherence) NMR spectra
of the reaction mixture at a flow rate of 8 $\mathrm{\mu L\,min^{-1}}$.
\cbstart
The conventional
pulse sequences were modified by replacing the initial \(\pi\)/2 pulse with a
\(\pi/4\ \)pulse; we refer to these experiments as ``PH-TOCSY'' (parahydrogen TOCSY) and
``PH-HMQC'' (parahydrogen HMQC).

A PH-TOCSY spectrum acquired in 20 min is shown in \fig{fig:PH-TOCSY-HMQC}a.
A \emph{thermal equilibrium} TOCSY spectrum of this compound would be
expected to
contain diagonal peaks connecting the identical nuclear spins in the two
acquisition dimensions, and off-diagonal peaks connecting \emph{J}-coupled
spins. In the PH-TOCSY experiment, the diagonal peaks only appear
for the two parahydrogen proton signals, because they are the only spins
significantly polarised in the indirect dimension. The other protons
are only polarised during the isotropic spin-mixing step of the pulse sequence,
and hence do not appear in the indirect dimension. These protons only produce
off-diagonal peaks, connecting them to the parahydrogen pair.
As shown in \fig{fig:PH-TOCSY-HMQC}b, the simulated spectrum
closely corresponds to the experimentally observed one.

A PH-HMQC spectrum acquired in 60 min is shown in \fig{fig:PH-TOCSY-HMQC}c.
It contains two peaks, linking the parahydrogen protons to the
\textsuperscript{13}C spins to which they have a direct
\textsuperscript{1}\emph{J}\textsubscript{CH} coupling.
An experiment of this kind, in which signals are
detected at full natural abundance of the \textsuperscript{13}C spins (about
1\%) in a 2.5~$\mu$L  detection volume, is only possible due to both the high
polarisation levels and stability of the system.
\cbstart
The sensitivity of the HMQC spectrum is limited by $t_1$ noise, due to imperfect
cancellation of the signals from molecular sites without \textsuperscript{13}C.
To some extent, this may stem from residual instability of the flow conditions,
as well as temperature fluctuations in the system. It may therefore be possible
to improve the sensitivity further by optimising the setup.
\cbend

The results in \fig{fig:PH-TOCSY-HMQC} show that the hyperpolarised spin order
can be spread to other
protons in the molecule by the application of the isotropic mixing
sequence MLEV-17
 \cite{levittSupercyclesBroadbandHeteronuclear1982,*baxMLEV17basedTwodimensionalHomonuclear1985}
 prior to 1D signal acquisition.
\cbend
 This simple trick
allows one to hyperpolarise any protons that are \emph{J}-coupled to the
parahydrogen pair, which makes the technique more general.

Much ongoing research in the field of hyperpolarisation is
motivated by in vivo applications, where hyperpolarised compounds
are used as magnetic
resonance imaging contrast agents \cite{hovener2018parahydrogen}.
Mostly, this involves transferring the
nuclear spin polarisation after hydrogenation to other nuclei
(\textsuperscript{13}C, \textsuperscript{15}N, \textsuperscript{31}P) with
lower magnetogyric ratios, where spin-lattice relaxation times are longer.
\cite{Goldman:2005bf,Goldman:2006cp,Reineri:2015he} Many of these approaches
use zero or very low magnetic fields for hydrogenation and polarisation
transfer. This has the advantage that near magnetic equivalence between the two
added protons is maintained through the reaction, leading to longer lifetimes
\cite{bhattacharya2007towards,chekmenev2008pasadena,
chekmenev2009hyperpolarized,shchepin2014parahydrogen,
Reineri:2015he,cavallari201813,ripka2018hyperpolarized,roy2018sabre}.
The present work opens a complementary strategy, in that the hydrogenation
is done at high field. Deleterious effects of relaxation are minimised by
the proximity of the site of hydrogenation to the point of use. Arguably,
this approach has advantages in the context of microfluidic systems, where
only small quantities of hyperpolarised agents are needed.
\cbstart
It may also be possible to adapt the present approach to the SABRE variant
of parahydrogen-induced hyperpolarisation\cite{Bordonali:2019jq}. In this context,
microfluidic recirculation may enable hyperpolarisation of very small total sample
volumes.
\cbend

\section{Conclusions}  The combination of a
highly efficient transmission-line NMR micro detector with para\-hyd\-ro\-gen-in\-duced
hyper\-polarisation leads to an un\-precedented sensitivity in inductively detected
NMR, with a mass limit of detection around 2.2~$\text{pmol}\,\sqrt{\mathrm{s}}$.
This corresponds to a concentration sensitivity of less than 1~$\mu \mathrm{M}\,\sqrt{\text{s}}$,
which, to our knowledge, has not previously been reached at the volume
scale of 2.5~$\mu$L.
This opens the perspective to be able to study chemical processes involving
low-abundance species in mass-limited samples. Obviously, such applications
require preparation of a hyperpolarised reactant. As the foregoing study shows,
the necessary chemistry can be integrated in a microfluidic system.
%To our knowledge, this
%is the best sensitivity reported for an inductively detected NMR signal
%to date.
It should be noted that we have used
parahydrogen enriched to 50\% (compared to 25\%
at thermal equilibrium); the sensitivity
could easily be boosted by a factor of three by using pure parahydrogen.
Microfluidic systems hold great potential in combination
with hyperpolarised NMR. All hyperpolarisation techniques require coordinated
manipulation of fluids and spin transformations. The results shown in the
foregoing demonstrate that in the case of parahydrogen-induced polarisation,
this can be assisted considerably by integrating some of the necessary chemical
steps on a microfluidic chip. Parahydrogen can be delivered to a reactive
solution through a PDMS membrane at sufficient rate to achieve significant
levels of hyperpolarisation; dissolution and transport of hydrogen in PDMS does
not appear to lead to significant ortho-para equilibration.
The highly stable continuous operation
of the PHIP@chip system allows quantitative studies
of the hydrogenation kinetics, and the relevant relaxation processes.
This is demonstrated by the dependence of the steady-state signal intensity on
flow rate and the recovery of the
hyperpolarised signal after saturation (\fig{fig:satrec-summary}).
Detailed kinetic and transport models, taking into account
the deprotection of the catalyst, hydrogen dissolution in the flowing
liquid, the hydrogenation reaction, and spin-lattice relaxation,
are currently being developed in our laboratory and will be reported
on a separate occasion.

The successful demonstration of PHIP on a chip opens important perspectives.
Conditions can be optimised for continued production of hyperpolarised metabolites,
which opens the possibility to conduct in-situ metabolic studies in microfluidic
cultures of cells, tissues, and organisms.
While the hyperpolarised compound used here, allyl acetate, is not
a metabolite per se, the production of hyperpolarised metabolic species
through PHIP has been demonstrated before \cite{cavallari201813,shchepin2014parahydrogen,reineri2015parahydrogen,Ripka:2018dc,Korchak:2018ga,hovener2018parahydrogen}.
Some metabolites, such as fumarate, can be
 generated directly by hydrogenation
of an unsaturated precursor \cite{Ripka:2018dc}. Aime et al. have proposed a more generally
applicable method \cite{reineri2015parahydrogen}, which relies on the metabolite bound to an alkyne sidearm
through an ester linkage. After hydrogenation, the polarisation is  transferred to a
 \textsuperscript{13}C nucleus in the metabolic moiety, and the sidearm is
 cleaved.  PHIP@chip opens the possiblity of implementing these additional
  production steps on the same chip.
\cbstart
  While previous demonstrations of sidearm hydrogenation have been carried
 out at low magnetic field, it may be possible to adapt recently developed
 efficient methods for heteronuclear polarisation transfer at
 high field\cite{eills2017singlet} to this purpose.
\cbend
In turn, this may enable integration
of the hyperpolarised metabolite generation with an on-chip culture
of cells or other biological systems.
Thanks to its stability,
the setup provides a convenient means to optimise pulse sequences
and reaction conditions for producing hyperpolarised targets.

\begin{acknowledgement}
%\section{Acknowledgements}
This work has been supported by the UK Engineering
and Physical Sciences Research Council (EPSRC) and Bruker UK Ltd.~in the the
form of a CASE conversion studentship for JE. Parts of the work presented here
were supported by the European Union Horizon 2020 Future and Emerging
Technologies programme (TISuMR project, grant number 737043).
\end{acknowledgement}

\begin{suppinfo}
   The supporting information contains a schematic of the PHIP@chip setup,
   a fully assigned spectrum (obtained with thermal H\textsubscript{2}),
   data on the ortho-para conversion of Hydrogen dissolved in PDMS, details
   of the finite element simulations, the saturation recovery data at all
   flow rates, and phase-sensitive plots of the simulated and experimental
   PH-TOCSY spectra.
\end{suppinfo}

% \section{Author contributions}
% J.E., W.H. and M.S. performed experiments. W.H. designed and manufactured the devices. J.E. adapted and implemented the pulse sequences used. M.R. contributed to the parahydrogen relaxation in PDMS study. M.H.L. provided guidance on pulse sequence phase cycling as well as simulation. M.U. performed simulations. M.U. conceived the project and the manuscript and SI were written by J.E., M.U., W.H. and M.H.L.

%\section{References}
%\bibliographystyle{achemso}
\bibliography{utz-collection,phip@chip-lit}

\end{document}